# QCT study of the vibrational and translational role in the $CN(v) + C_2H_6(v_1, v_2, v_5$ and $v_9)$ reactions

Joaquin Espinosa-Garcia, Cipriano Rangel, and Jose C. Corchado*

Área de Quimica Fisica and Instituto de Computación Científica Avanzada de Extremadura, Universidad de Extremadura, Badajoz (Spain),

e-mail: corchado@unex.es

## ABSTRACT

Quasi-classical trajectory (QCT) calculations were conducted on the newly developed full-dimensional potential energy surface, PES-2023, to analyse two critical aspects: the influence of vibrational versus translational energy in promoting reactivity, and the impact of vibrational excitation within similar vibrational modes. The former relates to Polanyi's rules, while the latter concerns mode selectivity. Initially, the investigation revealed that independent vibrational excitation by a single quantum of ethane's symmetric and asymmetric stretching modes (differing by only 15 $cm^{-1}$) yielded comparable dynamics, reaction cross-sections, HCN(v) vibrational product distributions, and scattering distributions. This observation dismisses any significant mode selectivity. Moreover, providing an equivalent amount of energy as translational energy (at total energies of 9.6 and 20.0 kcal $mol^{-1}$) gave rise to slightly lower reactivity compared to putting the same amount of energy as vibrational energy. This effect is more evident at low energies, presenting a counterintuitive scenario in an 'early transition state' reaction. These findings challenge the straightforward application of Polanyi's rules in polyatomic systems. Regarding CN(v) vibrational excitation, our calculations revealed that the reaction cross-section remains nearly unaffected by this vibrational excitation, suggesting the CN stretching mode is a spectator mode. The results were rationalized by considering several factors: the strong coupling between different vibrational modes, between vibrational modes and the reaction coordinate, and a significant vibrational energy redistribution within the ethane reactant before collision. This redistribution creates an unphysical energy flow, resulting in the loss of adiabaticity and vibrational memory before the reactants' collision. These theoretical findings require future confirmation through experimental or quantum mechanical theoretical studies, which are currently unavailable.

## 1. INTRODUCTION

The phenomena of vibrational excitation exert a significant influence on the outcomes of chemical reactivity, impacting both kinetics and dynamics. This influence is observed in reaction rates, pathways, and selectivity. In the atom-diatom scenario, Polanyi[1] pioneered this study by introducing the well-known "Polanyi's rules." Furthermore, Zare's group[2-4] as well as Crim's group[5-7] extended this study to polyatomic systems. This extension posed a considerable challenge, both experimentally and theoretically, due to the increased degrees of freedom.

According to Polanyi's rules, translational energy tends to dominate in "early" barrier reactions, mirroring reactant-like characteristics, while vibrational energy becomes more influential in "late" barrier reactions, resembling product-like characteristics. However, the application of these rules becomes notably more intricate in the case of "central" barrier reactions, which do not exhibit distinct reactant-like or product-like behaviour. Applying these rules to polyatomic reactions adds another layer of complexity due to the higher number of degrees of freedom and the potential for intramolecular vibrational energy redistribution.[8] A related concern is mode specificity, exploring the dynamics of different vibrational mode excitations within the reactants.

Within this multifaceted framework, we aim to examine the role of vibrational and translational energy in changing the dynamics of polyatomic reactions. Both forms of energy play pivotal roles in determining the reaction outcome. Vibrational energy rocks molecules along the potential energy surface, while translational energy directs the overall motion of reactants. The synergistic interplay between these energy forms can result in diverse reaction pathways, challenging traditional notions regarding activation barriers and unveiling the intriguing field of barrierless reactions.

In our research group, we have analyzed the influence of the translational/vibrational energy on the dynamics of the reaction in polyatomic systems, with conflicting results. In the OH + NH$_3$ reaction,[9] which exhibits an "early" barrier, translational energy proves more effective. Conversely, in the Cl($^2$P) + NH$_3$ reaction,[10] featuring a "late" barrier, the vibrational energy is more effective in driving the reaction. These finding represent a direct extension of Polanyi's rules to polyatomic reactions.

However, when dealing with "central" barriers, we have found contradictory results.[11, 12] For instance, in the O($^3$P) + CH$_4$ reaction, translational energy proves more effective than vibrational energy, whereas the opposite trend is observed in the H + C$_2$H$_6$ reaction. Even with polyatomic reactions displaying "early" barriers, we have encountered contradictory outcomes.[13][14] This is evident in the F($^2$P) + CH$_4$ reaction and its isotopomer analogue, F($^2$P) + CH$_2$D$_2$. The observed behaviour was explained[13] by the substantial energy transfer between vibrational modes in the deuterated reaction, nearly absent in the perprotio reaction, where modes essentially maintain their adiabaticity until interaction with fluorine. Wang and Czako,[15] using a time-dependent wavepacket method, also analysed the F($^2$P) + CH$_4$ → HF + CH$_3$ reaction, finding that translational energy is more effective than vibrational energy for energies below 0.38 eV; with the reverse being true for energies higher than 0.38 eV. This finding was confirmed few years later by Liu,[16] who studied methane bimolecular reactions, concluding that "on an equivalent amount of total energy, the initial translational energy is always more efficacious at lower collision energy in promoting the reaction rate than the reactant vibration, irrespective of the barrier location and the type of vibrational mode".

While extensive research has been devoted to the effect of vibrational excitation in methane reactions (see the comprehensive review in Ref. [16]), there remains a notable scarcity of both theoretical and experimental information regarding reactions involving ethane. To delve deeper into this realm, we began studying a series of ethane reactions,[12, 17] H + C$_2$H$_6$ and Cl($^2$P) + C$_2$H$_6$, both featuring a "central" barrier. While the chlorine reaction shows near-similar reactivity enhancement by vibrational and translational energies, our findings in the hydrogen reaction reveal that at lower collision energies, translational energy proves more effective than vibrational energy, while at higher collision energies, the opposite holds true—vibrational energy surpasses in effectiveness, validating Liu's previously drawn conclusions for methane reactions.[16] Additionally, the contributions by Czako's group[18-22] have significantly augmented our understanding of reactions involving ethane. In the present work we focus on analyzing the role of vibrational excitation in the polyatomic CN + C$_2$H$_6$ reaction, characterized by a distinct "early" barrier.

For the title reaction, CN($v$) + C$_2$H$_6$($v$) → HCN($v$) + C$_2$H$_5$($v$), involving 10 atoms and 24 degrees of freedom, an original analytical full-dimensional potential energy surface was recently developed in our group, named PES-2023.[23] It is essentially a valence-bond function with molecular mechanic terms, describing physically intuitive concepts, such as stretching, bending, and torsion about the C–C bond motions within the reactive system as it evolves from reactants to products. Utilizing this PES, a series of dynamics[23] and kinetics[24] studies were conducted for the vibrational ground states of both reactants. The dynamics analysis was carried out using the quasi-classical trajectory method, QCT, at various temperatures for comparison with experimental data. Our findings indicate that the majority of available energy was deposited as vibrational energy in the HCN product, and while the CH stretching mode exhibited an inverted vibrational population, the bending and CN stretching modes did not show inverted populations, consistent with experimental evidence. The kinetics analysis involved different approaches and was compared with experimental data across a broad temperature range of 25-1000 K. Our findings reveal that the obtained rate constants displayed a V-shaped temperature dependence, showing a minimum near 200 K, aligning with experimental evidence. The reasonable agreement of the dynamics and kinetics results with the experimental data instils confidence in the quality of the new PES-2023. This allows us to be optimistic about the outcomes of the present work, which, regrettably, as far as we are aware, cannot be compared with either experimental or theoretical information. Thus, this study marks the first exploration of the reactivity of vibrationally excited ethane and cyano radicals.

In this study, our focus lies in analysing the impact of vibrational excitation by one quantum number on specific modes: $v_1$ and $v_5$ ($CH_3$ stretching modes) and $v_2$ and $v_9$ ($CH_3$ deformation modes) in ethane, along with the CN stretching mode in the cyano radical. Our primary objective is twofold: first, to scrutinize mode selectivity, especially given the close proximity between the $v_1$ symmetric and $v_5$ asymmetric stretching modes (a mere 15 cm$^{-1}$ difference) and secondly, to investigate the influence of an equivalent amount of total energy, whether in the form of translational or vibrational energy, on the dynamics. The structure of this paper is organized as follows: Section 2 will provide a concise overview of the previously established PES-2023,[23] accompanied by an in-depth description of the vibrational modes in ethane along with their numerical values, and the computational details regarding the QCT calculations. In Section 3 the theoretical results for the CN($v$) + C$_2$H$_6$($v$) title reaction in its vibrational ground state and various vibrational excitations will be presented. Finally, Section 4 will summarize the key conclusions drawn from our study.

## 2. COMPUTATIONAL TOOLS

**A) Potential energy surface.** Theoretical study of a reactive process's dynamics requires knowing the potential energy surface (PES), which describes the nuclei's motion within the Born–Oppenheimer approximation. Recently, our group formulated an analytical full-dimensional PES for the title reaction.[23] This PES was fitted to high-level *ab initio* calculations at the explicitly correlated CCSD(T)-F12/aug-cc-pVTZ level. It is fundamentally based on a valence bond function enriched with molecular mechanic terms and switching functions to describe the evolution of the reactive system along the path from reactants to products, ensuring seamless transitions throughout the reaction's progression. In brief, it is formulated as the summation of five terms. The first term is related with stretching motions describing the six equivalent C-H bonds and the C-C bond in ethane. The second term describes the bending motions associated with the ethane-ethyl transformation in the reaction and is developed as the sum of harmonic terms. A third term describes the change from the pyramidal (-$CH_3$ group in ethane) to the planar (-$CH_2$ group in ethyl radical) geometry and is developed as a quadratic-quartic potential. The torsion around the C-C bond between the two methyl groups is described by a fourth term and, finally, a fifth function is included to describe the HCN product, CN stretching and HCN bending motions. These terms were described in detail in a previous paper,[23] and therefore will not be repeated here.

The title reaction exhibits high exothermicity (-22.20 kcal mol$^{-1}$), a very low barrier (+0.23 kcal mol$^{-1}$), and intermediate complexes in the entrance and exit channels, with the former showing minimal stabilization compared to the reactants at -0.27 kcal mol$^{-1}$. The transition state geometry resembles that of the reactants, classifying it as an "early" transition state.

**B) Vibrational modes in ethane.** Due to their significance in this study, the ethane harmonic vibrational frequencies obtained using PES-2023 are listed in Table 1, while Figure 1 illustrates the ethane vibrational modes analysed in the present study: $\nu_1$, $\nu_2$, $\nu_5$ and $\nu_9$. The modes $\nu_1$ (3011 cm$^{-1}$) and $\nu_5$ (2996 cm$^{-1}$) correspond to different $CH_3$ stretching motions -specifically symmetric and asymmetric, respectively. These modes were chosen to study mode selectivity in the title reaction due to their close proximity, with only a 15 cm$^{-1}$ difference. The modes $\nu_2$ (1428 cm$^{-1}$) and $\nu_9$ (1006 cm$^{-1}$) correspond to $CH_3$ bending modes and were selected to analyse the effect of bending excitations.

Additionally, we explored the first overtone of $\nu_2$ and second overtone of $\nu_9$ to investigate potential Fermi resonances,[25, 26] as the energies $2*\nu_2$ and $3*\nu_9$ closely align with those of the $\nu_1$ or $\nu_5$ modes.

To study the impact of an equivalent amount of energy as either translation or vibration on reactivity, a total energy of 9.6 kcal mol$^{-1}$ was chosen. Specifically, the following combinations of translational ($E_{coll}$) and vibrational ($E_{vib}$) energies were analysed (all units in kcal mol$^{-1}$):

$$CN(v=0) + C_2H_6\ (v=0) \qquad E_{coll} = 9.6 \qquad E_{total} = 9.6$$

$$CN(v=0) + C_2H_6\ (\nu_1=1) \qquad E_{coll} = 1.0 \qquad E_{total} = 9.6$$

$$CN(v=0) + C_2H_6\ (\nu_5=1) \qquad E_{coll} = 1.0 \qquad E_{total} = 9.6$$

$$CN(v=0) + C_2H_6\ (\nu_2=1) \qquad E_{coll} = 5.5 \qquad E_{total} = 9.6$$

$$CN(v=0) + C_2H_6\ (\nu_2=2) \qquad E_{coll} = 1.4 \qquad E_{total} = 9.6$$

$$CN(v=0) + C_2H_6\ (\nu_9=1) \qquad E_{coll} = 6.7 \qquad E_{total} = 9.6$$

$$CN(v=0) + C_2H_6\ (\nu_9=3) \qquad E_{coll} = 1.0 \qquad E_{total} = 9.6$$

$$CN(v=1) + C_2H_6\ (v=0) \qquad E_{coll} = 3.5 \qquad E_{total} = 9.6$$

Note that the frequencies of the $\nu_1$ and $\nu_5$ CH stretching modes differ by 15 cm$^{-1}$, which is equivalent to only 0.043 kcal mol$^{-1}$. Therefore, in practice, we considered the same excitation energy of 8.6 kcal mol$^{-1}$ in both vibrational modes. The same applies to the second overtone of $\nu_9$.

**C) Energy analysis.** The energy available to the products, denoted as $E_{avail.}$, comprises various contributions,

$$E_{avail.} = E_{coll} + E_{vib} + E_{rot} + \Delta H_R(0\ K) \qquad (1)$$

In all the cases, as previously mentioned, the sum of the first two terms, vibrational and collision energies in reactants, is 9.6 kcal mol$^{-1}$. The reactants' rotational energy, $E_{rot}$, was obtained from thermal sampling at room temperature which, averaging 1.0 kcal mol$^{-1}$. Lastly, the enthalpy of reaction at 0 K, $\Delta H_R(0\ K)$, is -26.0 kcal mol$^{-1}$. Thus, using the PES-2023 surface, the energy available to products is approximately 36.6 kcal mol$^{-1}$.

With such high energy available, correlated to the $C_2H_5$ co-product in its vibrational ground state, one should anticipate significant vibrational excitation in the vibrational modes of the HCN co-product. These modes are the CN stretching ($\nu_{CN}$, with a frequency of 2046 cm$^{-1}$), doubly degenerated bending, ($\nu_{bending}$, 723 cm$^{-1}$), and CH stretching modes, ($\nu_{CH}$, 3527 cm$^{-1}$). The energy allows up to six and three quanta in the CN and CH stretching modes, respectively, and up to 17 quanta in the bending mode, considering pure harmonic vibrational states. Additionally, numerous combination bands are feasible given the total available energy. Also, if the $C_2H_5$ co-product is observed in a vibrational excited state, the excitation of the correlated HCN product is expected to decrease.

**D) Computational details.** All dynamics calculations were performed using quasi-classical trajectory calculations (QCT), implemented in the VENUS code.[27, 28] The analytical PES-2023 surface[23] was employed for these calculations. The sum of vibrational and collisional reactant's energy is 9.6 kcal mol$^{-1}$. While the CN radical rotational energy remained fixed at the ground-state, the ethane rotational energy was selected via thermal sampling at 298 K, following a Boltzmann distribution. All trajectories, both ground-state and vibrationally excited, were initiated and terminated with an interatomic C-CN distance of 10.0 Å, ensuring negligible interaction in the reactant or product asymptotic regions. A step of 0.1 fs was employed to maintain energy conservation in each trajectory. For each reaction in Subsection 2B, a total of 100 000 trajectories were run. The maximum impact parameter, $b_{max}$, was determined by running 10 000 trajectories and progressively increasing the impact parameter, $b$, until no further reactive trajectories were identified. The $b_{max}$ value ranges between 4.0 to 5.0 Å, increasing with the vibrational excitation. Other initial conditions (impact parameter, vibrational phases, and spatial orientation) were chosen through Monte Carlo sampling.

QCT calculations rely on classical mechanics, presenting a significant limitation known as the zero-point energy (ZPE) violation problem. As a result, trajectories may result in both HCN and $C_2H_5$ products with a vibrational energy below their ZPEs, or trajectories where the vibrational energy of a specific mode drops below its ZPE, which is quantum mechanically prohibited. In both scenarios, a mixing of vibrational energy between modes occurs during the reaction evolution, leading to a phenomenon which is termed here as artificial vibrational energy redistribution (AVR). To address this issue, our previous paper[23] explored two approaches: a) the "All" approach, where all reactive trajectories were included in the final analysis and b) the "DZPE" (double ZPE) approach, considering only reactive trajectories with vibrational energy of each product -HCN and $C_2H_5$- above their respective ZPEs. Clearly, the DZPE approach is more physically plausible, albeit significantly reducing the number of reactive trajectories. Later, we will delve deeper into the approach utilized to gauge the AVR, specifically the loss of energy in each vibrational mode as the reaction progresses, a particularly intriguing aspect when studying vibrational excitation in polyatomic reactions.

After running the total number of trajectories, $N_T$, the reaction cross section, $\sigma_r$, can be determined from the number of reactive trajectories, $N_r$,

$$\sigma_r = \pi b_{max}^2 \frac{N_r}{N_T}. \qquad (2)$$

Due to the large number of trajectories run in each analysed reaction, the statistical error is minimal or negligible (< 2%), and thus, it is not shown in the Results Section.

Furthermore, rotational and vibrational energies of both products, $E^{rot}(C_2H_5)$, $E^{vib}(C_2H_5)$, $E^{rot}(HCN)$, and $E^{rot}(HCN)$, are computed, along with the relative translational energy, $E^{trans}$, between both products, and the scattering angle $\theta$ between the HCN product and the incident CN radical. The three actions for the HCN co-product, ($v_{CN}$, $v_{bending}$, $v_{CH}$). are obtained using our normal mode analysis (NMA) method.[29] In classical calculations, these actions are represented by real numbers. Various methods have been proposed to convert these actions into integer values or "quantize" them. The most cost-effective approach is the standard binning or histogram binning method (SB),[30] where each reactive trajectory contributes with a weight of unity to the final result. More computationally demanding, alternatives, like the Gaussian binning approach, GB,[31, 32] or the energy-based Gaussian binning method, 1GB,[33, 34] were considered in a previous study on the title reaction in its vibrational ground state.[23] Our conclusion was in favour of the SB alternative due to the significant population of numerous vibrational states. Here, we presume a similar scenario.

Before concluding this section, it's essential to comment on the AVR and NMA methods. Two in-house modifications[29] were made to the VENUS code. While the complete details are available in the original paper, we'll provide a concise overview here. For both methods, the Cartesian coordinates and momenta at the final step of each trajectory were projected onto the normal mode space, enabling the computation of kinetic and potential energies in each normal mode. The total energy is obtained from the summation of these terms and, using a harmonic description of the energy levels, actions can be calculated. It is important to distinguish between the two methodologies: in the AVR estimation, where the vibrational redistribution in the reactant channel before the reactants' collision is obtained, the CN and $C_2H_6$ reactants serve as references; whereas in the NMA approach, which focuses on the products' vibrational distributions, the HCN and $C_2H_5$ products are used as the reference structures.

## 3. RESULTS AND DISCUSSION

**3.1. Influence of vibrational excitation on reactivity: translational versus vibrational energy.** The QCT reaction cross-section ratios for the various analysed vibrational excitations -CN($v$=1) and $C_2H_6$($\nu_1$, $\nu_2$, $\nu_5$, $\nu_9$)- in relation to the reactants vibrational ground-states, CN($v$=0) + $C_2H_6$($v$=0), are presented in Table 2. These ratios were calculated at a total energy of 9.6 kcal mol$^{-1}$, using the All and DZPE trajectory counting methods. We begin by analysing the behaviour of the CN vibrational excitation. The reaction cross-section remains practically unaffected by the vibrational excitation as observed with both counting methods, with the maximum error bar considered at ±0.1. The CN mode acts as a spectator mode, showing minimal impact on the reaction. In contrast, ethane vibrational excitation slightly enhances reactivity, by factors ranging 1.0-2.4. Notably it is observed that the DZPE method tends to overestimate these ratios compared to the All approach. This behaviour arises from the exclusion of a greater number of reactive trajectories in the ground-state calculations within the DZPE counting method due to the ZPE violation problem, consequently increasing the $\sigma_v/\sigma_{gs}$ ratio. Hence, caution is advisable when interpreting the results obtained through the DZPE counting method. Thus, vibrational excitation appears to be slightly more effective in steering the reaction compared to an equal amount of energy as translation. This initially appears counterintuitive within an "early" barrier reaction scenario.

Next, we examine the excitations near the energies of possible Fermi resonances. The close energy proximity between the $2\nu_2$ (or $3\nu_9$) and $\nu_1$ (or $\nu_5$) states, with differences ≤150 cm$^{-1}$ in the $2\nu_2$ case and < 22 cm$^{-1}$ in the $3\nu_9$ case, suggest the potential for a mixing of vibrational wavefunctions, displaying Fermi resonances. However, it's important to note that in our classical calculations we cannot account for this quantum effect, making this analysis speculative within our study. Our results (Table 2) indicate that the $2\nu_2$ and $3\nu_9$ vibrational excitations enhance reactivity in comparison to the $\nu_2$ and $\nu_9$ excitations, showing similarities to the effects observed with the $\nu_1$ or $\nu_5$ excitations. To summarize, while our calculations are classical, our results prompt speculation regarding the possible existence of Fermi resonances in this reaction. However, due to the lack of experimental data, these theoretical results await conformation or refutation from future experimental or quantum mechanical studies.

To provide comprehensive insight, additional QCT calculations were conducted at a higher total energy of 20.0 kcal mol$^{-1}$ to explore the impact of vibrational excitation on reactivity. The reaction cross-section ratios in comparison to the ground states of the reactants are also shown in Table 2. When examining the results with the reactants in their vibrational ground states, it is evident that the reaction cross-section increases with the collision energy, from 4.192 to 7.941 Å$^2$ when employing the All counting method, a trend consistent with the expected behaviour. We observed a comparable enhancement upon vibrational excitation of the reactants, noted using both the All and DZPE counting methods. Furthermore, the cross-section ratios prove that the impact of ethane's vibrational excitation on reactivity remains virtually unchanged with varying total energy, indicating independence from the counting method employed. Consequently, at this energy an equal input of energy as translation or vibration yields a comparable effect on reactivity. Once more, this result presents a counterintuitive scenario within an "early transition state" reaction.

**3.2. Influence of vibrational excitation on reactivity: mode selectivity.** One of the primary objectives of this study was to examine whether different vibrational mode excitations in ethane are equally effective in promoting the reaction, constituting a case of mode selectivity. The $v_5$ and $v_1$ CH stretching modes in ethane differ by only 15 cm$^{-1}$, and they were independently excited by one quantum. At first glance one might anticipate different reactivity, since the $v_5$ stretch mode, being coupled to the reaction coordinate, is the reactive mode, while the $v_1$ appears to act as a spectator mode. However, as shown in Table 2, both modes, C$_2$H$_6$($v_1$ = 1) and C$_2$H$_6$($v_5$ = 1), exhibit a similar increase in reactivity with respect to the vibrational ground state. At a total energy of 9.6 kcal mol$^{-1}$, this increase is 1.557 and 1.646, respectively, and at 20.0 kcal mol$^{-1}$, it stands at 1.000 and 1.015. Hence, in this case we can dismiss mode selectivity for the title reaction.

**3.3. Understanding the role of vibrational excitation on reactivity.** To comprehend the influence of vibrational excitation on reactivity and shed light on the unexpected behaviour discussed in Sections 3.1 and 3.2, four qualitative approaches were employed. The first approach involved analysing the coupling terms, $B_{i,F}(s)$. These terms depict the coupling between mode $i$ and the reaction coordinate $F$,[35] serving as a measure of the energy flow between them. They are defined as

$$B_{i,F}(s) = -\sum_{l\gamma=1}^{N} \frac{dg_{l\gamma}(s)}{ds} c_{l\gamma}^{i}(s) \tag{3}$$

and from them, the reaction path curvature, $\kappa(s)$, can be obtained,

$$\kappa(s) = \left\{\sum [B_{i,F}(s)]^2\right\}^{\frac{1}{2}}. \tag{4}$$

Secondly, within the framework of the Hamiltonian of the reaction path,[35] we computed the coupling between modes, $B_{i,j}(s)$, which are defined as

$$B_{i,j}(s) = \sum_{l\gamma=1}^{N} \frac{dc_{l\gamma}^{i}(s)}{ds} c_{l\gamma}^{j}(s). \tag{5}$$

Note that in Eqs. 3 and 5, $c_{l\gamma}^{i}(s)$ is the $l\gamma$ component of the eigenvector for mode $i$ and $g_{l\gamma}(s)$ is the $l\gamma$ component of the normalized gradient vector.

Thirdly, we employed the sudden vector projection (SVP) model,[36-38] which also incorporates rotational and translational coupling with the reaction coordinate as well.

Finally, we analysed energy redistribution AVR,[29] which quantifies the gain or loss of classical vibrational energy in each vibrational mode as reaction progresses.

Regarding the first approach, concerning the coupling between vibrational modes and the reaction coordinate, Figure 2 illustrates the curvature of the reaction path plotted against s, revealing four distinct peaks. Peak A, situated prior to the saddle point, originates from the strong coupling of the reaction coordinate to several modes: the CH stretching mode ($v_7$, doubly degenerated) and the $v_5$ reactive mode. Additionally, it involves contributions from the doubly degenerate CH$_3$ deformation ($v_{11}$), and the CH$_3$ deformations $v_2$ and $v_6$. The remaining peaks are located after the saddle point. Peak B appears due the coupling of the reaction coordinate to the CH stretching mode $v_7$ (doubly degenerated) and CH$_3$ deformations ($v_2$, $v_6$ and $v_{11}$). Peak C shows coupling of the reaction coordinate with the CH$_3$ rocking ($v_9$, doubly degenerated) and, finally, peak D represents its coupling with the $v_5$ reactive mode, the CN stretching mode and the $v_6$ CH$_3$ deformation mode. This analysis of the coupling between the reaction coordinate and vibrational modes reveals strong coupling among many vibrational modes and the reaction coordinate, suggesting a potential flow of energy between them. Consequently, one might anticipate enhancement in reactivity with the vibrational excitation of these modes. Surprisingly, despite the strong coupling of the $v_5$ CH asymmetric stretching mode to the reaction coordinate and the lack of coupling for the $v_1$ CH symmetric stretching mode, we do not observe a significant difference in the impact on the reactivity upon excitation of either mode. We will revisit this point later.

In terms of coupling between vibrational modes, Figure 3 displays the $B_{i,j}(s)$ terms. Overall, the figure depicts medium to strong coupling existing among numerous modes, facilitating energy transfer between them and leading to loss of adiabaticy in the process. Specifically, the $v_1$ mode shows coupling with other $CH_3$ stretching modes, while the $v_5$ reactive mode exhibits coupling with $CH_3$ stretching and bending modes. Additionally, the $v_2$ and $v_9$ bending modes primarily couple with the CN stretching mode.

A simpler picture of the process is attained from the SVP analysis output (Table 3). In this analysis, all $C_2H_6$ and CN vibrational modes exhibit minimal or negligible coupling with the reaction coordinate. Thus, the SVP analysis suggests that translational energy is more effective in promoting reactivity compared to vibration. This finding aligns with a direct application of Polanyi's rules to polyatomic systems but contradicts both the QCT results obtained (refer to Table 2) and the conclusions drawn from the analysis of the coupling terms.

Lastly, we examine a recognized limitation of classical QCT results: the classical energy transfer between vibrational modes during the reaction progression. In practical terms, in reactions involving vibrational excitation in polyatomic reactants, this translates to the loss of initial vibrational energy distribution memory. Consequently, a mixing of vibrational excitation occurs even before the reaction initiates, before there is noticeable interaction energy between the two reactants. The results of the AVR analysis are depicted in Figure 4. We observed an energy flow between modes even in the ethane vibrational ground-state (Panel A). Hence, all $CH_3$ stretching modes experience slight deactivation, leading to the transfer of this energy to $CH_3$ deformation modes. As a result, adiabaticity is not maintained before the collision of the reactants. When the $v_1$ and $v_5$ $CH_3$ stretching modes are excited independently by one quantum (Panels B and C), they rapidly deactivate (less than 0.2 ps) transferring their vibrational energy to several $CH_3$ bending modes ($v_4$, $v_6$, $v_8$, $v_{11}$ and $v_{12}$). Upon excitation of the $CH_3$ bending modes $v_2$ and $\nu_9$ by one quantum (Panels D and E), deactivation occurs at a slower rate, about 1.0 ps, with this energy flowing into the $\nu_6$, $\nu_8$ and $\nu_{12}$ CH$_3$ deformation modes. Finally, the vibrational excitation of the CN reactant is examined (Panel F). Both, vibrational ground-state and excitation by one quantum, maintain the adiabaticy prior to the reactants' collision, i.e., they preserve the memory until collision with the other reactant. Hence, we anticipate that our conclusions regarding the $\nu_2$ and $\nu_9$ bending excitation and CN stretching excitation remain valid. However, our findings concerning the excitation of the stretching modes might be somewhat influenced by the AVR. Thus, one should approach the conclusions drawn from classical trajectories on the effect of excitation in these modes with caution. Furthermore, this observation could shed light on the discrepancy between our analysis of the $B_{i,F}(s)$ terms and the QCT results. The energy deposited in these modes appears to flow to other modes, making it challenging to discern the specific effect of vibrational excitation on either mode.

To further explore this matter concerning the impact of translational/vibrational energy on polyatomic system reactivity, it is worth noting the similar behavior previously observed[13, 39,] in studies of the F + CH$_4$ and F + CH$_2$D$_2$ reactions both on the PES-2006 surface.[40] Despite both reactions being strongly exothermic with an "early transition state," contrasting behavior was noted depending on isotopic substitutions: in the F + CH$_4$ reaction, translation energy proved more effective than vibrational energy, whereas the opposite trend was observed in the F + CH$_2$D$_2$ reaction. These divergent results were interpreted based on the AVR behaviour in the entry channels. In the perprotio case, the system maintains adiabatic character for a longer duration, whereas in the CH$_2$D$_2$ molecule, energy mode transfer occurs faster, resulting in the loss of adiabatic character.

In conclusion, these coupling effects and artificial energy transfers, inherent to the classical nature of QCT calculations, pose challenges in analyzing the impact of vibrational excitation and the balance between translational and vibrational energy in the reactivity enhancement of this reaction. Consequently, more reliable outcomes are anticipated from our calculations on the $\nu_2$ and $\nu_9$ bending excitation, whereas stretching excitation might be affected by a lack of adiabaticity between modes, potentially undermining the reliability of the results. We hope that these conflicting findings will stimulate future experimental or quantum mechanics research in this domain.

**3.4 Role of vibrational excitation on product energy partitioning.** With respect to vibrational excitation's impact on product energy partitioning (Table 4), at a collision energy of 9.6 kcal mol$^{-1}$, f$_v$, f$_r$, and f$_{transl}$, the fractions of available energy allocated to vibration, rotation, and translation of the products, are obtained. Primarily, the bulk of available energy is channelled into HCN product vibrational energy (50-60%) with a minimal transfer to translational energy (10-20%). Upon exciting the CN reactant by one quantum, there's an increase in the HCN product's vibrational fraction compared to the vibrational ground-state (∼16%), mostly at the expense of translational energy (∼8%). This shy increase indicates that the vibrational energy is not maintained through the reaction path, signalling a loss of selectivity. Regarding ethane, excitation in the lowest modes, $v_2$ and $v_9$, is not maintained along the reaction path, suggesting their role as spectator modes. In the same way, independent excitation of $v_1$ and $v_5$ modes yield a similar picture, negating mode selectivity. Consequently, the additional vibrational energy primarily contributes to HCN and C$_2$H$_5$ product vibrational energies (approximately 10% increase in each product), yet this extra excitation doesn't endure the reaction. This analysis of the average property aligns with previously reported findings.

**3.5 Role of vibrational excitation on the HCN($v_1$,$v_2$,$v_3$) product vibrational distribution.** Before delving into this analysis, it is worth highlighting that the QCT/NMA results at a total energy of 9.6 kcal mol$^{-1}$ reveal over 300 populated HCN($v_1$,$v_2$,$v_3$) vibrational states, with 150 states exhibiting a population >0.1%. The most populated state, HCN(1,0,1), holds a mere 5% population, which aligns with findings from our previous work on the CN + C$_2$H$_6$ vibrational ground state reaction at 298 K.[23] These distributions underscore the challenge experimentalists fact in measuring these low populations.

In the preceding subsection, we noted that a significant portion of available energy allocated to HCN product vibrational energy (50-60%). Here, we explore how this vibrational energy is distributed among the HCN vibrational modes and the influence of reactant vibrational excitation on this property. Starting from the CN($v$=0) + C$_2$H$_6$($v$=0) ground state, the HCN($v_1$,$v_2$,$v_3$) exhibits only 1% population in its vibrational ground-state (0,0,0). The CN stretch, the bending and the CH stretch modes appeared excited, reaching up to $v_1$=3, $v_2$ =14 and $v_3$=3. Figure 5 illustrates the effects of the reactant vibrational excitation on these mode populations. Concerning the CH stretch mode in HCN, calculations for CN($v$=0) + C$_2$H$_6$($v$=0) showed that 40% of the population remains in its ground state. However, upon exciting either the $v_1$ or $v_5$ stretching modes of ethane to their first excited state, the population in the CH($v$=0) ground state diminishes to 20%.

Conversely, the $v=2$ and $v=3$ populations exhibit notable increments due to this mode's excitation. This suggests that the original excitation of the $v_1$ and $v_5$ modes in ethane does not persist through the reaction. Additionally, both $v_1$ and $v_5$ modes yield similar outcomes, indicating a loss of selectivity. Further analysis on the causes of this behaviour was previously discussed in subsection 3.3.

Upon analysing the CN($v$) population in HCN, we observe that the initial excitation in the CN($v$) reactant is retained in the products, consistent with the expectations from the AVR analysis (Figure 4). Finally, the HCN bending population remains relatively unchanged despite the original excitation of the $v_2$ and $v_9$ bending modes in ethane.

**3.6 Role of vibrational excitation on the HCN(v) product scattering distribution.** To investigate the influence of reactant vibrational excitation, Figure 6 displays the scattering of the HCN product with respect to the incident CN reactant at total energy of 9.6 kcal mol$^{-1}$. Note that 18 bins within the range 0-180º at 10º intervals were used for the product angular distribution, represented by the differential cross section (DCS) in Å$^2$ sr$^{-1}$, obtained from an expansion in Legendre polynomials.[41] When considering reactants in their vibrational ground-states, CN(v=0) + C$_2$H$_6$(v=0), a predominantly sideways scattering distribution emerged, peaking around 60°. However, upon introducing vibrationally-excited stretching states in ethane ($v_1$ or $v_5$ = 1), the scattering angles changed to a nearly isotropic behaviour. This shift can be attributed to the enlarged impact parameter due to excitation, widening the 'cone of acceptance' and thereby broadening the range of scattered angles in the products. Interestingly, vibrational excitation of the ethane bending modes ($v_2$ or $v_9$ = 1) barely altered the scattering distribution compared to the ground-state, as in both cases similar impact parameters are found. In the CN($v=1$) + C$_2$H$_6$($v=0$) reaction, a slightly more backward distribution was observed, despite presenting similar impact parameters to those of the vibrational ground-state.

It is crucial to exercise caution in interpreting these results for two reasons. Firstly, we believe that all angular distributions are influenced by the presence of a complex in the product channel, which complicates analysis and induces disorientation, leading trajectories to forget the memory of the initial impact orientation. Secondly, the y-axis scale used in Figure 6 is limited to values between 0.2 and 0.6. Such a small range of values amplifies the perception of the relative forward-backward-sideways behaviours.

However, in general terms, they are quite isotropic. Again, since no experimental data are available for comparison, these QCT angular distributions serve as predictive studies, awaiting confirmation or refutation in future theoretical or experimental studies.

## 4. CONCLUSIONS

To analyse the impact of vibrational excitation in the CN($v$) + C$_2$H$_6$($v_1$, $v_2$, $v_5$, $v_9$) gas-phase reaction on the reactivity and dynamics, exhaustive QCT calculations were carried out on the full-dimensional PES-2023 surface at two total energies, 9.6 and 20.0 kcal mol$^{-1}$.

1. The vibrational excitation of the $v_1$, $v_2$, $v_5$ and $v_9$ modes in ethane results in only marginal reactivity increases compared to the reaction involving reactants in their vibrational ground state, with enhancements ranging from 1.0 to 2.4, notably lower at higher energies. Notably, the most considerable enhancement is observed with stretching mode excitation, where our AVR computations suggest a lack of adiabaticity, due to substantial mode-mode coupling and energy flow between modes occurring before reactant collision. This minimal reactivity boost aligns with predictions from the SVP method but contradicts the findings from the analysis of coupling between vibrational modes and reaction coordinate, where significant curvature is observed before the saddle point. Overall, all the models consistently indicate minor or negligible reaction enhancements resulting from vibrational excitation.

2. Vibrational excitation in the CN($v$) reactant shows minimal impact on reactivity, suggesting its role a spectator mode.

3. When independently excited by one quantum, the symmetric and asymmetric stretching modes of ethane, differing by only 15 cm$^{-1}$, displayed analogous dynamic properties: similar reaction cross sections, HCN($v$) product vibrational distributions, and product scattering distributions. Consequently, the observed resemblance in their behaviours rules out the possibility of mode selectivity.

4. By using equal amounts of energy as vibration or translation, we found that vibrational energy increases reactivity only slightly more effectively than translation, which, a priori, represents a counterintuitive example taking into account that this reaction is an example of "early transition state". These results disagree with the predictions of the SVP analysis but are in agreement with our analysis of the coupling between vibrational modes and the reaction coordinate.

5. We found in our QCT calculations strong coupling between different vibrational modes and strong artificial vibrational redistribution in the reactants before collision. Therefore, part of the energy originally deposited in a vibrational mode can flow to other vibrational modes before reaction occurs, producing a loss of effectiveness in enhancing reactivity and a loss of adiabaticity and of memory prior to the reactants' collision. Therefore, these factors, which are concomitant to the complexity of classical dynamics on polyatomic systems with many degrees of freedom, can mask the analysis.

6. Finally, it is important note that neither experimental not quantum mechanics theoretical results are available for comparison, and thus the present results acquire a predictive character.

**REFERENCES**

(1) Polanyi, J. C. Concepts in Reaction Dynamics. *Acc Chem Res* **1972**, *5* (5), 161–168. https://doi.org/10.1021/ar50053a001.

(2) Bechtel, H. A.; Camden, J. P.; Brown, D. J. A.; Zare, R. N. Comparing the Dynamical Effects of Symmetric and Antisymmetric Stretch Excitation of Methane in the Cl+CH4 Reaction. *J Chem Phys* **2004**, *120* (11), 5096–5103. https://doi.org/10.1063/1.1647533.

(3) Camden, J. P.; Bechtel, H. A.; Brown, D. J. A.; Zare, R. N. Effects of C–H Stretch Excitation on the H+CH4 Reaction. *J Chem Phys* **2005**, *123* (13). https://doi.org/10.1063/1.2034507.

(4) Kim, Z. H.; Bechtel, H. A.; Camden, J. P.; Zare, R. N. Effect of Bending and Torsional Mode Excitation on the Reaction Cl+CH4→HCl+CH3. *J Chem Phys* **2005**, *122* (8). https://doi.org/10.1063/1.1844295.

(5) Yoon, S.; Henton, S.; Zivkovic, A. N.; Crim, F. F. The Relative Reactivity of the Stretch–Bend Combination Vibrations of CH4 in the Cl (2P3/2)+CH4 Reaction. *J Chem Phys* **2002**, *116* (24), 10744–10752. https://doi.org/10.1063/1.1476318.

(6) Yoon, S.; Holiday, R. J.; Crim, F. F. Control of Bimolecular Reactions: Bond-Selected Reaction of Vibrationally Excited CH3D with Cl (2P3/2). *J Chem Phys* **2003**, *119* (9), 4755–4761. https://doi.org/10.1063/1.1591176.

(7) Yoon, S.; Holiday, R. J.; Sibert, E. L.; Crim, F. F. The Relative Reactivity of CH3D Molecules with Excited Symmetric and Antisymmetric Stretching Vibrations. *J Chem Phys* **2003**, *119* (18), 9568–9575. https://doi.org/10.1063/1.1615755.

(8) Nesbitt, D. J.; Field, R. W. Vibrational Energy Flow in Highly Excited Molecules: Role of Intramolecular Vibrational Redistribution. *J Phys Chem* **1996**, *100* (31), 12735–12756. https://doi.org/10.1021/jp960698w.

(9) Monge-Palacios, M.; Espinosa-Garcia, J. Role of Vibrational and Translational Energy in the OH + NH$_3$ Reaction: A Quasi-Classical Trajectory Study. *J Phys Chem A* **2013**, *117* (24), 5042–5051. https://doi.org/10.1021/jp403571y.

(10) Monge-Palacios, M.; Corchado, J. C.; Espinosa-Garcia, J. Quasi-Classical Trajectory Study of the Role of Vibrational and Translational Energy in the Cl(2P) + NH3 Reaction. *Physical Chemistry Chemical Physics* **2012**, *14* (20), 7497. https://doi.org/10.1039/c2cp40786h.

(11) Espinosa-Garcia, J. Quasi-Classical Trajectory Study of the Vibrational and Translational Effects on the O($^3$P) + CD$_4$ Reaction. *J Phys Chem A* **2014**, *118* (20), 3572–3579. https://doi.org/10.1021/jp502414e.

(12) Espinosa-Garcia, J.; Calle-Cancho, J.; Corchado, J. C. QCT Study of the Vibrational and Translational Role in the H + C2H6(N1, N2, N5, N7, N9 and N10) Reactions. *Theor Chem Acc* **2019**, *138* (10), 116. https://doi.org/10.1007/s00214-019-2504-4.

(13) Espinosa-García, J. Vibrational versus Translational Energies in the F+CH4 Reaction: A Comparison with the F+CH2D2 Reaction Using Quasi-Classical Trajectory Methods. *Chem Phys Lett* **2010**, *488* (4–6), 153–157. https://doi.org/10.1016/j.cplett.2010.02.035.

(14) Espinosa-García, J. Quasiclassical Trajectory Calculations Analyzing the Role of Vibrational and Translational Energy in the F+CH2D2 Reaction. *J Chem Phys* **2009**, *130* (5). https://doi.org/10.1063/1.3069632.

(15) Wang, D.; Czakó, G. Quantum Dynamics Study of the F + CH$_4$ → HF + CH$_3$ Reaction on an Ab Initio Potential Energy Surface. *J Phys Chem A* **2013**, *117* (32), 7124–7130. https://doi.org/10.1021/jp4005778.

(16) Liu, K. Vibrational Control of Bimolecular Reactions with Methane by Mode, Bond, and Stereo Selectivity. *Annu Rev Phys Chem* **2016**, *67* (1), 91–111. https://doi.org/10.1146/annurev-physchem-040215-112522.

(17) Corchado, J. C.; Chamorro, M. G.; Rangel, C.; Espinosa-Garcia, J. State-to-State Dynamics of the Cl(2P) + C2H6(N5, N1 = 0, 1) → HCl(v', J') + C2H5 Hydrogen Abstraction Reactions. *Theor Chem Acc* **2019**, *138* (2), 26. https://doi.org/10.1007/s00214-019-2416-3.

(18) Papp, D.; Li, J.; Guo, H.; Czakó, G. Vibrational Mode-Specificity in the Dynamics of the Cl + C2H6 → HCl + C2H5 Reaction. *J Chem Phys* **2021**, *155* (11). https://doi.org/10.1063/5.0062677.

(19) Papp, D.; Czakó, G. Vibrational Mode-Specific Dynamics of the F(2P3/2) **+** C2H6 → HF **+** C2H5 Reaction. *J Chem Phys* **2021**, *155* (15). https://doi.org/10.1063/5.0069658.

(20) Gruber, B.; Tajti, V.; Czakó, G. Vibrational Mode-Specific Dynamics of the OH + C$_2$H$_6$ Reaction. *J Phys Chem A* **2023**, *127* (35), 7364–7372. https://doi.org/10.1021/acs.jpca.3c04328.

(21) Yin, C.; Czakó, G. Vibrational Mode-Specific Quasi-Classical Trajectory Studies for the Two-Channel HI + C$_2$H$_5$ Reaction. *Physical Chemistry Chemical Physics* **2023**, *25* (14), 9944–9951. https://doi.org/10.1039/D2CP05993B.

(22) Yin, C.; Czakó, G. Theoretical Vibrational Mode-Specific Dynamics Studies for the HBr + $C_2H_5$ Reaction. *Physical Chemistry Chemical Physics* **2023**, *25* (4), 3083–3091. https://doi.org/10.1039/D2CP05334A.

(23) Espinosa-Garcia, J.; Rangel, C. The CN(X 2Σ+) + C2H6 Reaction: Dynamics Study Based on an Analytical Full-Dimensional Potential Energy Surface. *J Chem Phys* **2023**, *159* (12). https://doi.org/10.1063/5.0172489.

(24) Espinosa-Garcia, J.; Bhowmick, S. Kinetic Study of the CN + C2H6 Hydrogen Abstraction Reaction Based on an Analytical Potential Energy Surface. **2023**. https://doi.org/10.48550/arXiv.2312.12117.

(25) Fermi, E. Über Den Ramaneffekt Des Kohlendioxyds. *Zeitschrift für Physik* **1931**, *71*, 250–259.

(26) Amat, G.; Pimbert, M. On Fermi Resonance in Carbon Dioxide. *J Mol Spectrosc* **1965**, *16* (2), 278–290. https://doi.org/10.1016/0022-2852(65)90123-2.

(27) Hase, W. L.; Duchovic, R. J.; Hu, X. Y.; Komornicki, A.; Lim, K. F.; hong Lu, D.; Peslherbe, G. H.; Swamy, K. N.; Linde, S. R. Vande; Varandas, A. J. C.; Wang, H.; Wolf, R. J. VENUS96: A General Chemical Dynamics Computer Program; 1996.

(28) Hu, X.; Hase, W. L.; Pirraglia, T. Vectorization of the General Monte Carlo Classical Trajectory Program VENUS. *J Comput Chem* **1991**, *12* (8), 1014–1024. https://doi.org/10.1002/jcc.540120814.

(29) Corchado, J. C.; Espinosa-Garcia, J. Product Vibrational Distributions in Polyatomic Species Based on Quasiclassical Trajectory Calculations. *Physical Chemistry Chemical Physics* **2009**, *11* (43), 10157. https://doi.org/10.1039/b912948k.

(30) Truhlar, D. G.; Muckerman, J. T. Reactive Scattering Cross Sections III: Quasiclassical and Semiclassical Methods. In *Atom - Molecule Collision Theory*; Springer US: Boston, MA, 1979; pp 505–566. https://doi.org/10.1007/978-1-4613-2913-8_16.

(31) Bonnet, L. The Method of Gaussian Weighted Trajectories. III. An Adiabaticity Correction Proposal. *J Chem Phys* **2008**, *128* (4). https://doi.org/10.1063/1.2827134.

(32) Bonnet, L. Classical Dynamics of Chemical Reactions in a Quantum Spirit. *Int Rev Phys Chem* **2013**, *32* (2), 171–228. https://doi.org/10.1080/0144235X.2012.752905.

(33) Czakó, G.; Bowman, J. M. Quasiclassical Trajectory Calculations of Correlated Product Distributions for the F+CHD3(V1=,1) Reactions Using an *Ab Initio* Potential Energy Surface. *J Chem Phys* **2009**, *131* (24). https://doi.org/10.1063/1.3276633.

(34) Bonnet, L.; Espinosa-García, J. The Method of Gaussian Weighted Trajectories. V. On the 1GB Procedure for Polyatomic Processes. *J Chem Phys* **2010**, *133* (16). https://doi.org/10.1063/1.3481781.

(35) Miller, W. H.; Handy, N. C.; Adams, J. E. Reaction Path Hamiltonian for Polyatomic Molecules. *J Chem Phys* **1980**, *72* (1), 99–112. https://doi.org/10.1063/1.438959.

(36) Jiang, B.; Guo, H. Relative Efficacy of Vibrational vs. Translational Excitation in Promoting Atom-Diatom Reactivity: Rigorous Examination of Polanyi's Rules and

(36)     Proposition of Sudden Vector Projection (SVP) Model. *J Chem Phys* **2013**, *138* (23). https://doi.org/10.1063/1.4810007.

(37)     Li, J.; Guo, H. Mode Specificity and Product Energy Disposal in Unimolecular Reactions: Insights from the Sudden Vector Projection Model. *J Phys Chem A* **2014**, *118* (13), 2419–2425. https://doi.org/10.1021/jp501255t.

(38)     Jiang, B.; Guo, H. Mode Specificity, Bond Selectivity, and Product Energy Disposal in X + $CH_4$/$CHD_3$ (X=H, F, O($^3$P), Cl, and OH) Hydrogen Abstraction Reactions: Perspective from Sudden Vector Projection Model. *Journal of the Chinese Chemical Society* **2014**, *61* (8), 847–859. https://doi.org/10.1002/jccs.201400158.

(39)     Espinosa-García, J. Quasiclassical Trajectory Calculations Analyzing the Role of Vibrational and Translational Energy in the F+CH2D2 Reaction. *J Chem Phys* **2009**, *130* (5). https://doi.org/10.1063/1.3069632.

(40)     Espinosa-García, J.; Bravo, J. L.; Rangel, C. New Analytical Potential Energy Surface for the F($^2$P) + $CH_4$ Hydrogen Abstraction Reaction: Kinetics and Dynamics. *J Phys Chem A* **2007**, *111* (14), 2761–2771. https://doi.org/10.1021/jp0688759.

(41)     Truhlar, D. G.; Blais, N. C. Legendre Moment Method for Calculating Differential Scattering Cross Sections from Classical Trajectories with Monte Carlo Initial Conditions. *J Chem Phys* **1977**, *67* (4), 1532–1539. https://doi.org/10.1063/1.435057.

**Table 1**. Harmonic vibrational frequencies for ethane using PES-2023 surface.

| Mode | Description | Frequency[a] |
|---|---|---|
| $\nu_1$ | CH$_3$ stretch | 3011 |
| $\nu_2$ | CH$_3$ deform | 1428 |
| $\nu_3$ | CC stretch | 1189 |
| $\nu_4$ | Torsion | 292 |
| $\nu_5$ | CH$_3$ stretch | 2996 |
| $\nu_6$ | CH$_3$ deform | 1405 |
| $\nu_7$ | CH$_3$ stretch | 3013 (d) |
| $\nu_8$ | CH$_3$ deform | 1522(d) |
| $\nu_9$ | CH$_3$ rock | 1006(d) |
| $\nu_{10}$ | CH$_3$ stretch | 3031(d) |
| $\nu_{11}$ | CH$_3$ deform | 1543(d) |
| $\nu_{12}$ | CH$_3$ rock | 825(d) |

a) Frequency in cm$^{-1}$, where (d) means doubly degenerate. Note that the CN radical presents a single vibrational frequency of 2114 cm$^{-1}$.

**Table 2.** Reaction cross-section ratios for different CN(v) + C$_2$H$_6$(v) vibrational excitations[a] with respect to the CN(v=0) + C$_2$H$_6$(v=0) vibrational ground-states (gs), using the PES-2023 surface.

| | $\sigma_v/\sigma_{gs}$ | |
|---|---|---|
| | QCT-All | DZPE |
| **Ethane modes. Total energy of 9.6 kcal mol$^{-1}$** | | |
| $v_1$ | 1.557 | 2.295 |
| $v_5$ | 1.646 | 2.345 |
| $v_2$ | 1.007 | 1.254 |
| 2*$v_2$ | 1.509 | 2.173 |
| $v_9$ | 1.414 | 1.788 |
| 3*$v_9$ | 1.705 | 2.663 |
| **Cyano Mode** | | |
| $v$ | 1.079 | 0.945 |
| **Ethane modes. Total energy of 20.0 kcal mol$^{-1}$** | | |
| $v_1$ | 1.000 | 1.586 |
| $v_5$ | 1.015 | 1.558 |
| $v_2$ | 0.997 | 1.314 |
| $v_9$ | 0.969 | 1.219 |

a) All modes are excited by a quantum, unless otherwise indicated.

**Table 3**. SVP projections of vibration, translation and rotation reactant modes using the PES-2023 surface.

| Species | Mode | Frequency[a] | SVP |
|---|---|---|---|
| $C_2H_6$ | $CH_3$ stret | 3011 | 0.017 |
| | $CH_3$ deform | 1428 | 0.006 |
| | CC stret | 1189 | 0.012 |
| | Torsion | 292 | 0.002 |
| | $CH_3$ stret | 2996 | 0.017 |
| | $CH_3$ deform | 1405 | 0.001 |
| | $CH_3$ stret | 3013(d) | 0.020 |
| | $CH_3$ deform | 1522(d) | 0.005 |
| | $CH_3$ rock | 1006(d) | 0.009 |
| | $CH_3$ stret | 3031(d) | 0.010 |
| | $CH_3$ deform | 1543(d) | 0.005 |
| | $CH_3$ rock | 825(d) | 0.003 |
| CN | stret | 2114 | 0.004 |
| $C_2H_6 \cdots CN$ | Translation | | 0.926 |
| | Rotation | | 0.362 |

a) Frequency in $cm^{-1}$, where (d) means doubly degenerate

**Table 4.** Product energy partitioning (%)[a] for different $C_2H_6(v)$ vibrational excitations using the PRS-2023 surface and a total energy of 9.6 kcal mol$^{-1}$.

|  | $f_v(C_2H_5)$ | $f_r(C_2H_5)$ | $f_v(HCN)$ | $f_r(HCN)$ | $f_t$ |
|---|---|---|---|---|---|
| **$C_2H_6(v=0)$+ $CN(v=0)$** | 10 | 10 | 50 | 11 | 19 |
| $v_1$ | 19 | 6 | 61 | 6 | 8 |
| $v_5$ | 18 | 5 | 62 | 7 | 8 |
| $v_2$ | 15 | 8 | 55 | 9 | 13 |
| $v_9$ | 13 | 9 | 53 | 10 | 15 |
| **$C_2H_6(v=0)$+ $CN(v=1)$** | 10 | 7 | 66 | 6 | 11 |

a) $f_v$, $f_r$ and $f_t$ mean, respectively, fraction of energy as vibration, rotation and translation. The DZPE counting method is used and all vibrations excited by one quantum.

**FIGURE CAPTION**

**Figure 1**. Normal-mode vector representation for the $\nu_1$, $\nu_2$, $\nu_5$ and $\nu_9$ ethane vibrational modes, obtained using the PES-2023 surface.

**Figure 2**. Reaction path curvature, $\kappa(s)$, as a function of reaction coordinate, $s$.

**Figure 3.** Coriolis-like $B_{i,i'}(s)$ coupling terms (a.u.) along the reaction path. For a clearer exposition, small or negligible coupling terms are not represented. The normal modes are numbered from highest to lowest vibrational frequencies in both X and Y axes, and the colours show the intensity of the coupling.

**Figure 4.** Average energy (kcal mol$^{-1}$) of each normal mode in the reactants as a function of time. The ethane vibrational modes are listed in Table 1.

**Figure 5**. Effects of the reactants' vibrational excitations on the QCT/PES-2023 vibrational distributions at 9.6 kcal mol$^{-1}$ for the HCN($v_1,v_2,v_3$) states where all co-product states were considered.

**Figure 6.** HCN product angular distributions (with respect to the CN reactant) for the CN ($v$) + C$_2$H$_6$($v$) reactions, normalized by the factor ($2\pi/\sigma_r$).

FIGURE 1

**Mode 1: 3011 cm$^{-1}$**

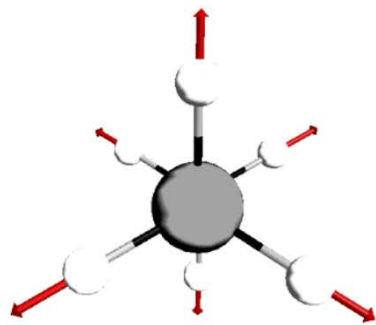

**Mode 5: 2996 cm$^{-1}$**

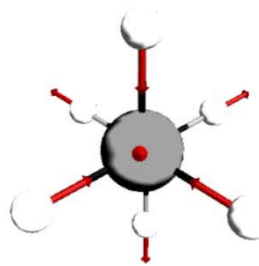

**Mode 2: 1428 cm$^{-1}$**

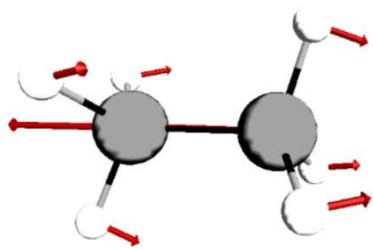

**Mode 9: 1006 cm$^{-1}$**

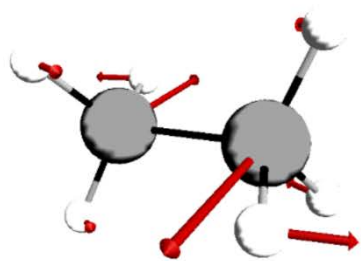

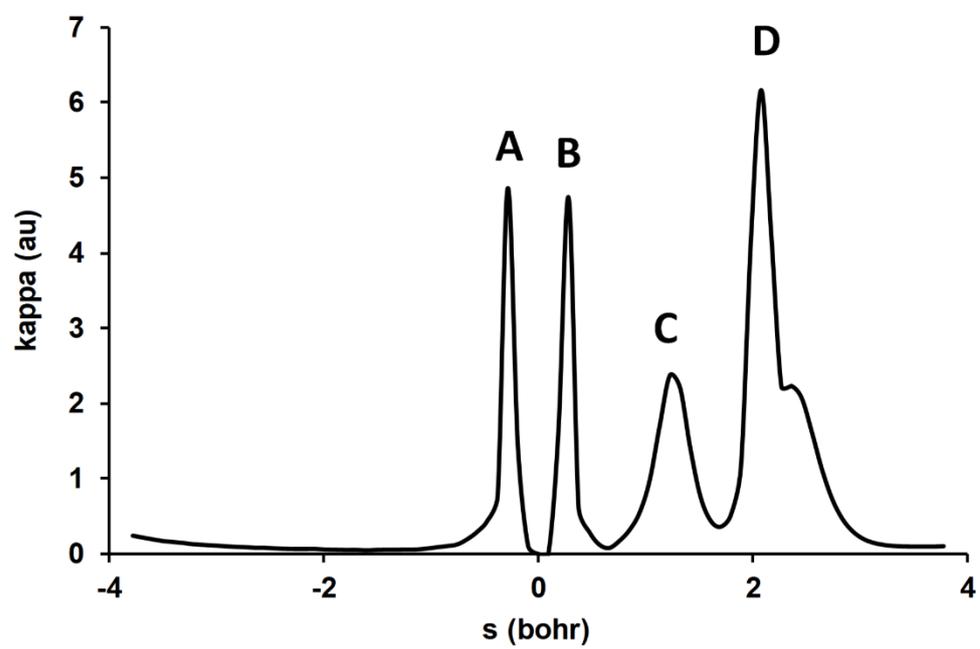

**FIGURE 2**

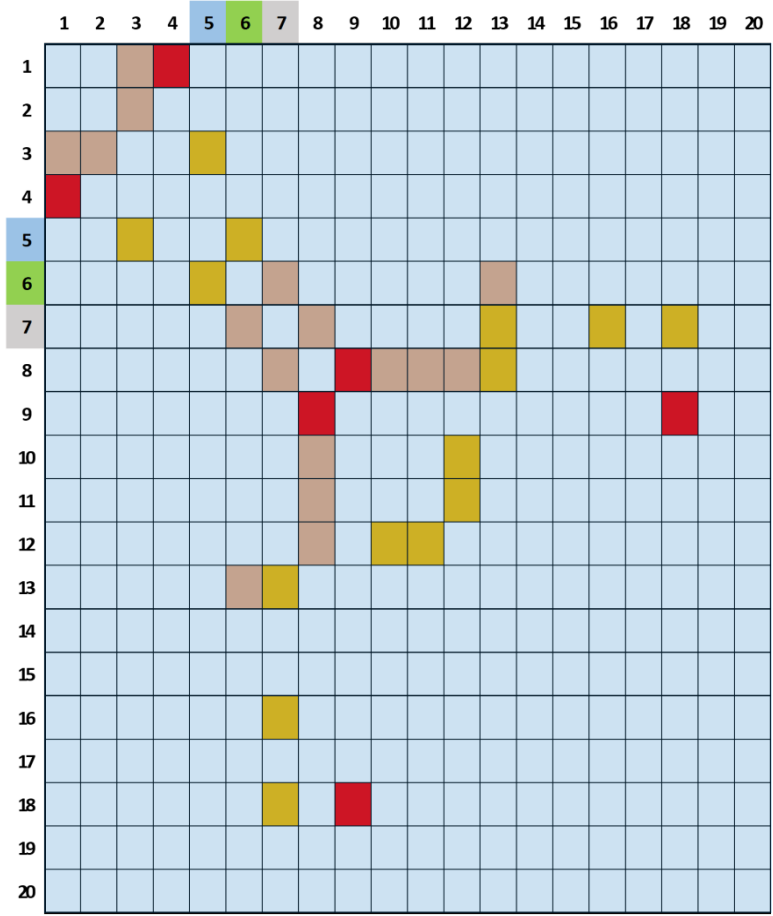

**FIGURE 3**

### A) Ethane Ground state

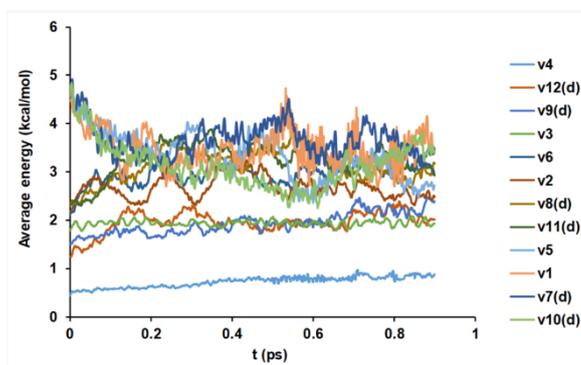

### B) $v_1$ mode excited

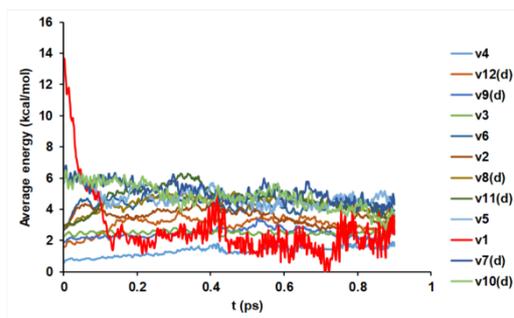

### C) $v_5$ mode excited

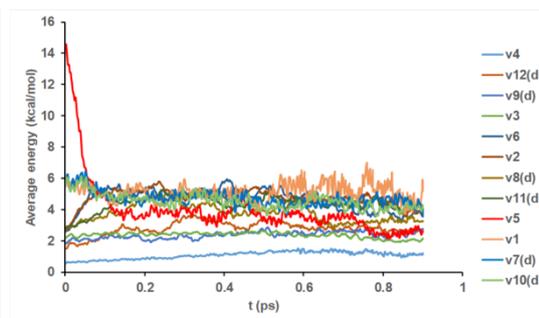

### D) $v_2$ mode excited

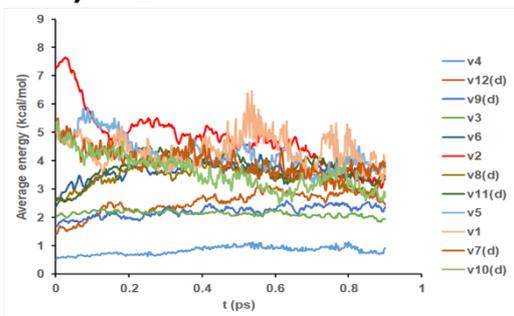

### E) $v_9$ mode excited

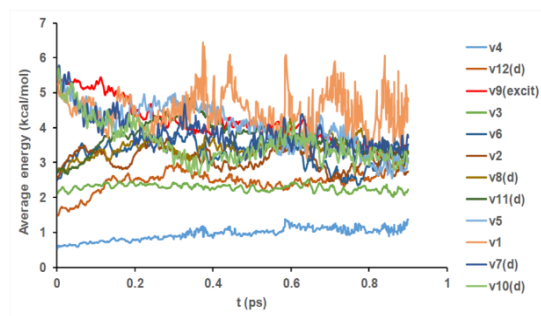

### F) CN (v=0,1)

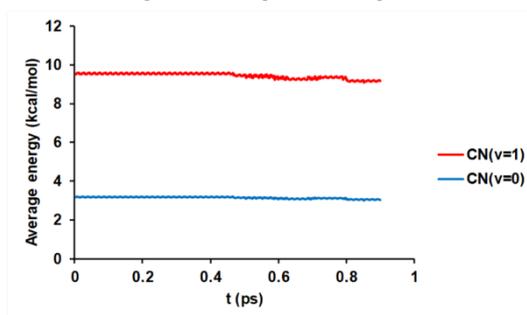

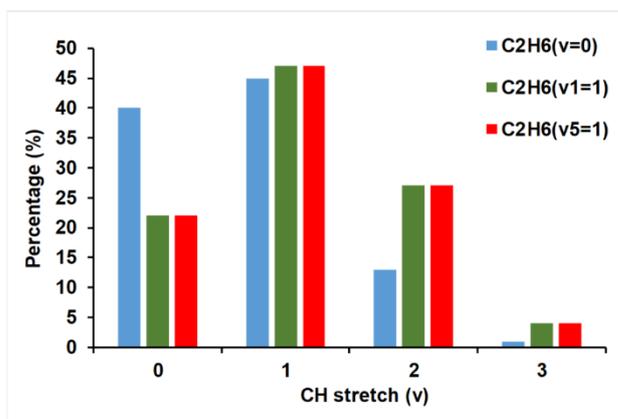

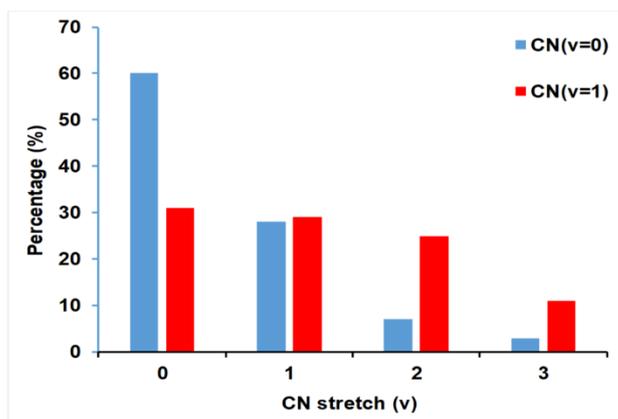

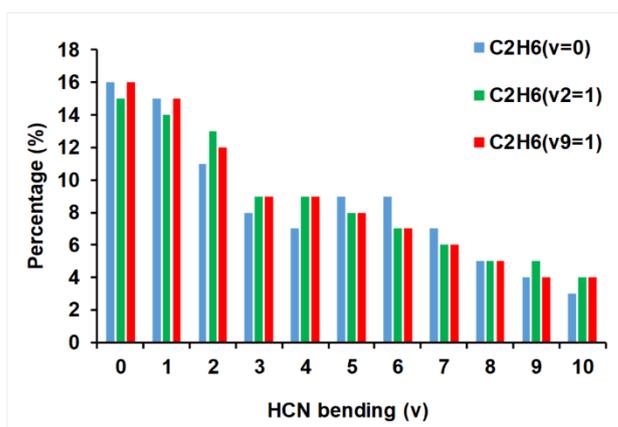

**FIGURE 6**

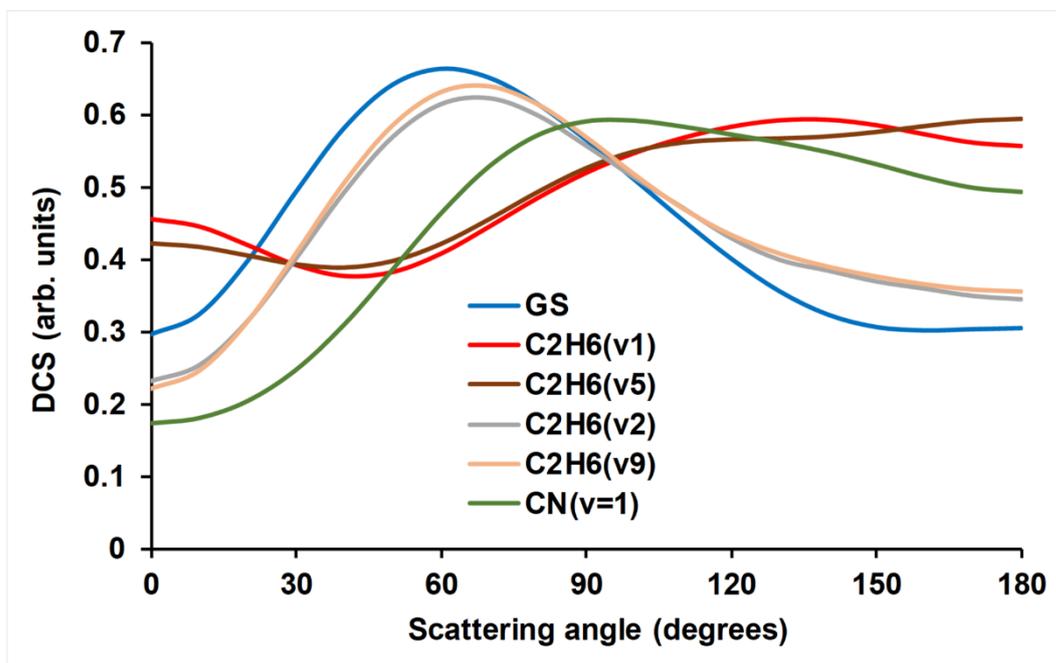